# Multi-state Chiral Switching Through Adiabaticity Control in Encircling Exceptional Points

Aodong Li[1,5], Jian Wang[1], Andrea Alù[2,3] and Lin Chen[1,4,*]

[1] Wuhan National Laboratory for Optoelectronics and School of Optical and Electronic Information, Huazhong University of Science and Technology, Wuhan 430074, China

[2] Photonics Initiative, Advanced Science Research Center, City University of New York, New York, NY 10031, USA

[3] Physics Program, Graduate Center, City University of New York, New York, NY 10016, USA

[4] Shenzhen Huazhong University of Science and Technology Research Institute, Shenzhen 518063, China

[5] School of Science and Engineering, The Chinese University of Hong Kong, Shenzhen, 518172, China

Email: chen.lin@mail.hust.edu.cn

* Corresponding author.

## Abstract

Dynamic encircling of exceptional points has attracted significant interest in recent years, as it can facilitate chiral transmission selectivity due to a nontrivial eigenstate evolution. Recently, multi-state systems have been explored, associated with more complex topologies supporting a larger number of exceptional points, but chiral switching among multiple eigenstates has remained elusive in experiments. Here, we overcome this challenge by dividing the eigenstate space into multiple subspaces by controlling the adiabaticity. The eigenstates in different subspaces can evolve without crosstalk, and chiral switching occurs as the eigenstates within each subspace are subject to a non-adiabatic transition while they encircle exceptional points. We experimentally demonstrate this phenomenon by reporting chiral switching for two groups of optical modes at telecom wavelengths in a four-state optical system, and theoretically demonstrate that our approach can be extended to higher-order systems. Our findings pave new avenues for studying chiral dynamics based on exceptional-point physics in multi-state systems, and offer opportunities to develop multiplexed photonic devices.

## Introduction

Tailored non-Hermitian systems have been shown to support exceptional points (EPs), which have attracted a great deal of interest across physical disciplines, including optics[1-4], acoustics[5-7], electronics[8-10], thermodynamics[11] and quantum mechanics[12-14]. As spectral singularities associated with self-intersecting Riemann surfaces (RSs), EPs give rise to intriguing optical phenomena, such as single-mode lasing[15-17], enhanced sensing[18-22] and topological mode engineering[23-25], which are of fundamental importance for a variety of breakthrough technologies.

Recent work on non-Hermitian systems has revealed that the eigenstate evolution can undergo nonadiabatic transitions (NATs), not available in Hermitian systems. NATs can arise while dynamically encircling EPs, resulting in counterintuitive and technologically relevant chiral responses, e.g., the output state is independent of the input state, but solely determined by the encircling direction[26-36]. Up to now, most EP-encircling studies have focused on two-state systems, leading to chiral switching between a pair of eigenstates. Quite recently, interest on EP encircling has shifted from two-state to multi-state systems, associated with more complex topologies, holding the promise for new opportunities for manipulating multiple eigenstates[37-39]. However, any input state is bound to trigger the lowest-loss eigenstate during the evolution, due to lack of a mechanism for selectively governing non-adiabatic phenomena among multiple eigenstates. Consequently, the output is locked to the single eigenstate with lowest loss. If there exists multiple lowest-loss eigenstates, the final state will be a superposition of them[39]. For this reason, to the best of our knowledge chiral switching with multiple eigenstates has remained elusive in experiments.

In this paper, we overcome this challenge by grouping eigenstates into multiple subspaces, controlling the non-adiabaticity degree as we evolve among them. The eigenstates supported within different invariant direct sum subspaces are tailored to have zero non-adiabaticity degree, and hence they are not subject to mutual excitation during the Hamiltonian evolution. The eigenstates within each subspace can be triggered due to non-zero non-adiabaticity, and chiral switching between them occurs due to NAT as we encircle EPs. Based on this paradigm, we experimentally demonstrate chiral switching in a four-state optical system at telecom wavelengths, and theoretically extend our approach to multi-state chiral switching in higher-order dimensions.

## Results

### Principle of adiabaticity control in encircling exceptional points

When the Hamiltonian $H(\tau)$ slowly changes over time $\tau$ in a Hermitian system, the system remains in the same eigenstate throughout the evolution, based on the adiabatic theorem[40-42]. However, this theorem is not applicable for evolutions in non-Hermitian systems. For an *N*-order non-Hermitian Hamiltonian system, with $|\varphi_m\rangle$ and $|\varphi_n\rangle$ being two arbitrary eigenstates (or called right eigenstates), $|\varphi_m\rangle$ being the initial eigenstate and $|\varphi_n\rangle$ the eigenstate with lowest loss ($m, n \in [0, N-1]$), $|\varphi_n\rangle$ can be excited during the evolution, since adiabaticity is not stringently fulfilled, and $|\varphi_n\rangle$ becomes the final output state due to a nontrivial NAT[31,39]. The adiabaticity parameter is defined as $\xi_{m,n} = |(\hbar \langle \tilde{\varphi}_m | \dot{\varphi}_n \rangle)/(E_m - E_n)|$, where $\hbar$ is the reduced Planck constant, $\langle \tilde{\varphi}_m|$ is the left eigenstate with $H^\dagger |\tilde{\varphi}_m\rangle = E_m^* |\tilde{\varphi}_m\rangle$, $|\dot{\varphi}_n\rangle = \partial |\varphi_n\rangle / \partial \tau$, $E_m$ and $E_n$ are the corresponding eigenvalues for $|\varphi_m\rangle$ and $|\varphi_n\rangle$, respectively, and the dot above the function denotes the time derivative. To prevent the excitation of $|\varphi_n\rangle$ during the evolution, $\xi_{m,n}$ should be zero. The output can be mixed with multiple eigenstates having simultaneously the lowest loss, if the non-adiabaticity degree is not purposely controlled[39].

The eigenstate at $\tau$, $|\varphi_n(\tau)\rangle$, can be expressed as the superposition of all eigenstates at the initial time $\tau_0$, $|\varphi_n(\tau)\rangle = \sum_{j=0}^{j=N-1} c_{j,n}(\tau) |\varphi_j(\tau_0)\rangle$, with $c_{j,n}(\tau)$ being the expansion coefficients. The numerator of $\xi_{m,n}$ can be then expressed as

$$\langle\tilde{\varphi}_m(\tau_0)|\dot{\varphi}_n(\tau_0)\rangle = \langle\tilde{\varphi}_m(\tau_0)|\lim_{\Delta\tau\to 0}\frac{|\varphi_n(\tau_0+\Delta\tau)\rangle - |\varphi_n(\tau_0)\rangle}{\Delta\tau}$$

$$= \lim_{\Delta\tau\to 0}\frac{\sum_{j=0}^{j=N-1}c_{j,n}(\tau_0+\Delta\tau)\langle\tilde{\varphi}_m(\tau_0)|\varphi_j(\tau_0)\rangle - \langle\tilde{\varphi}_m(\tau_0)|\varphi_n(\tau_0)\rangle}{\Delta\tau} \quad (1)$$

Because the left and right eigenstates of $H$ are bi-orthogonal, i.e., $\langle\tilde{\varphi}_m|\varphi_m\rangle = 1$ and $\langle\tilde{\varphi}_m|\varphi_n\rangle = 0$ for $m \neq n$ [43]. Eq. (1) can be simplified as $\langle\tilde{\varphi}_m(\tau_0)|\dot{\varphi}_n(\tau_0)\rangle = \dot{c}_{m,n}(\tau_0)$. We can let $c_{m,n}(\tau_0) = 0$ to yield $\xi_{m,n} = 0$. By using $P(\tau) = [|\varphi_0(\tau)\rangle \ |\varphi_1(\tau)\rangle \ \cdots \ |\varphi_{N-1}(\tau)\rangle]$ to represent all eigenstates of $H(\tau)$, the transformation relation of eigenstates at $\tau$ and $\tau_0$ can be described as $P(\tau) = P(\tau_0)C$, where $C$ is the transformation matrix. If the eigenstate space can be divided into $K$ direct sum subspaces as $P = P_1 \oplus P_2 \oplus \cdots \oplus P_K$ ($P_k$ is an invariant subspace under the transformation, $k = 1, 2, \ldots, K$), the transformation relation can be rewritten as

$$[P_1(\tau) \ P_2(\tau) \ \cdots \ P_K(\tau)] = [P_1(\tau_0) \ P_2(\tau_0) \ \cdots \ P_K(\tau_0)]\begin{bmatrix} C_1 & 0 & & \\ 0 & C_2 & & \\ & & \ddots & 0 \\ & & 0 & C_K \end{bmatrix} \quad (2)$$

with $P_k(\tau) = P_k(\tau_0)C_k$. Therefore, any two eigenstates belonging to different $P_k$ satisfy $\xi_{m,n} = 0$, i.e., an initial eigenstate in one subspace does not trigger any other eigenstate in a different subspace during the evolution.

Interestingly, a Hamiltonian with a double-symmetric matrix form can group all eigenstates into two subspaces. The double-symmetric form indicates the $N \times N$ Hamiltonian $H$ is symmetric, with respect to its main diagonal and skew diagonal elements, simultaneously. The eigenstates in one subspace are symmetric, and the eigenstates in the other subspace are anti-symmetric. The eigenstates in different subspaces cannot excite each other during the evolution (see Supplementary Note 1 for a detailed demonstration). A four-coupled optical waveguide system can then be designed to support a double-symmetric Hamiltonian with

$$H = \begin{bmatrix} \beta - i\gamma & \kappa & & \\ \kappa & -\beta & \kappa & \\ & \kappa & -\beta & \kappa \\ & & \kappa & \beta - i\gamma \end{bmatrix} \quad (3)$$

where $\beta$, $\gamma$ and $\kappa$ represent the detuning, relative loss rate, and coupling strength, respectively. The eigenvalue spectrum of $H$ forms two sets of self-intersecting RSs, RS1 and RS2, in the parameter space $(\beta/\kappa, \gamma/\kappa)$, as schematically presented in Fig.1. Two eigenstates are located in the upper surfaces, with an EP at $(0.5, 2)$, and the other two eigenstates are located in the lower surfaces, with an EP at $(-0.5, 2)$. Figure 1 shows the dynamic trajectory of the Hamiltonian for clockwise (CW) (Fig. 1a) and anti-clockwise (ACW) loops (Fig. 1b) around the two EPs.

The state evolution obeys a Schrödinger-type equation, $i\partial/\partial\tau |\varphi\rangle = H |\varphi\rangle$. At the starting point $(0,0)$ of the trajectory, the four eigenstates from top to bottom are $|\varphi_0\rangle = [a,b,b,a]^T$, $|\varphi_1\rangle = [b,a,-a,-b]^T$, $|\varphi_2\rangle = [b,-a,-a,b]^T$, and $|\varphi_3\rangle = [a,-b,b,-a]^T$ ( $a = \sqrt{5-\sqrt{5}}/2\sqrt{5}$ and $b = \sqrt{5+\sqrt{5}}/2\sqrt{5}$ ), respectively, in which $|\varphi_0\rangle$, $|\varphi_2\rangle$ are symmetric, and $|\varphi_1\rangle$, $|\varphi_3\rangle$ are anti-symmetric. An eigenstate with eigenvalue with larger imaginary part suffers lower loss.

For CW encircling, $|\varphi_0\rangle$ ($|\varphi_1\rangle$) evolves along the yellow (green) line on the red surface with low loss. The system ends at $|\varphi_2\rangle$ ($|\varphi_3\rangle$), caused by the self-intersecting property of the RS. The two evolution trajectories are located on two different RSs, as indicated by the right panels in Fig. 1a. The two evolution processes do not interfere with each other, as the symmetric and anti-symmetric eigenstates belong to different subspaces. For the ACW encircling direction, as shown in Fig. 1b, $|\varphi_0\rangle$ ($|\varphi_1\rangle$) initially evolves on the RS with high loss and switches to the eigenstates with low loss on the upper (lower) red surfaces due to a NAT. Two eigenstates in each subspace are associated with one RS. Consequently, the system state will jump to the red RS1, rather than to the red RS2, when $|\varphi_0\rangle$ is injected. Meanwhile, the system will jump to the red RS2 when $|\varphi_1\rangle$ is injected. In other words, the adopted grouping method ensures that only one eigenstate with low loss is dominant after the NAT process. Finally, $|\varphi_0\rangle$ and $|\varphi_1\rangle$ return to their initial state. Overall, the system can independently enable chiral transmission for the two eigenstate groups. For the encircling trajectories on RS1 (RS2), the output state is locked to $|\varphi_2\rangle$ ($|\varphi_3\rangle$) and $|\varphi_0\rangle$ ($|\varphi_1\rangle$) for CW and ACW handedness, respectively (see Supplementary Note 2 for the evolution paths with the injected states being of $|\varphi_2\rangle$ and $|\varphi_3\rangle$) For CW and ACW evolution, we have numerically calculated the dynamic evolution along the encircling trajectories presented in Fig. 1, verifying chiral switching for each group of eigenstates (see Supplementary Note 3).

**Experimental verification of multi-state chiral switching**

We map our theory onto a realistic geometry consisting of four-coupled optical waveguides, associated with a double-symmetric Hamiltonian, as schematically shown in Fig. 2a. The following design strategies have been employed to map the Hamiltonian parameters in Eq. (3) onto the device structure. First, we used sufficiently large gap separation between adjacent waveguides to introduce weak coupling, guaranteeing a symmetric Hamiltonian along the main diagonal. Then, the four waveguides are symmetrically positioned with respect to the center (denoted by green plane), making sure that the Hamiltonian is symmetric along the anti-diagonal. Finally, the gap separations between adjacent waveguides are kept the same.

For the first and fourth waveguides, a metallic chromium strip of 20-nm thickness was used to introduce a position-dependent absorption loss. The chromium strip is tapered on both sides, with the purpose of preventing reflections caused by abrupt structural change. Based on coupled-mode theory, the relationship between the structural parameters and the Hamiltonian parameters can be then established[44]. The Hamiltonian parameters $\beta$ and $\gamma$ are proportional to $w_1 - w_2$ and $w_m$, respectively, and $\kappa$ is inversely proportional to $d$ [45]. The detailed relationship can be found in Fig. S3 of Supplementary Note 4. The evolution time $\tau$ is mapped to the propagation distance $z$, and the CW and ACW loops correspond to light propagation along the positive and negative directions with respect to the $z$ axis, respectively. The EP-encircling evolution trajectory is schematically shown in Fig. 2b. At the two ports of the device - positions A and D in Fig. 2a - all four waveguides have the same width $w_1 = w_2$. The four-coupled waveguides support four eigenmodes, i.e., $TE_0$, $TE_1$, $TE_2$, and $TE_3$ modes, corresponding to the four eigenstates $|\varphi_0\rangle$, $|\varphi_1\rangle$, $|\varphi_2\rangle$, and $|\varphi_3\rangle$. These four modes are divided into two groups, i.e., the even modes $TE_0$ and $TE_2$, and the old modes $TE_1$ and $TE_3$. The four coupled waveguides for chiral mode switching are linked to bus waveguides using two branches on both ends, which are used for our experimental demonstration.

Figures 3a-3d show the electric field evolution during propagation in a four-coupled silicon waveguides system, simulated with the Finite-Difference Time-Domain (FDTD) method (see Supplementary Note 5 for the detailed geometrical parameters of the four-coupled silicon waveguides). When the $TE_0$ mode enters from the left port, it propagates to the two middle waveguides from A to B, as shown in the first row of Fig. 3a. The mode evolution from B and C experiences almost no energy loss, and finally evolves to $TE_2$ at D. For the $TE_0$ mode excited from the right port, as shown in the second row of Fig. 3a, the system evolves to the side waveguides from D to C, but quickly attenuates from C and B, due to absorption in the Cr layer. A portion of the energy leaks due to non-adiabatic evolution into the middle waveguides. This branch of mode evolution suffers no loss, and hence it becomes the dominant mode, associated with a NAT, and finally the energy exits at A as $TE_0$. As a summary, $TE_2$ and $TE_0$ modes are produced when $TE_2$ enters from the left and right ports, respectively (Fig. 3b). In other words, the output mode is locked to $TE_2$ ($TE_0$) for the left-side (right-side) input, regardless of $TE_0$ or $TE_2$ mode injection. The other group of eigenmodes, i.e., $TE_1$ and $TE_3$ modes, show a similar evolution process as $TE_0$ and $TE_2$ modes (Figs. 3c and 3d). The output mode is locked to $TE_3$ ($TE_1$) for the left-side (right-side) input, regardless

of a $TE_1$ or $TE_3$ mode injection. Benefitting from selectively governing the non-adiabaticity degree, during the entire evolution one group of eigenmodes, i.e., $TE_0$ and $TE_2$ modes, will not excite the other group of eigenmodes, i.e., $TE_1$ and $TE_3$ modes, and vice versa.

A scanning electron microscope (SEM) image of the four-coupled silicon waveguides in one of the fabricated samples is shown in Fig. 4a (See Method and Supplementary Note 6 for the fabrication details). Figures 4b-4e show the measured transmittance for different mode inputs within 1540-1560 nm. The measurement details can be found in Supplementary Note 7. The output is dominated by $TE_2$ ($TE_0$) mode when $TE_0$ (Fig. 4b) and $TE_2$ (Fig. 4c) modes excite the left (right) port, and is dominated by $TE_3$ ($TE_1$) mode when $TE_1$ (Fig. 4d) and $TE_3$ (Fig. 4e) modes input from left (right) port. These measured results clearly demonstrate chiral switching for two groups of optical eigenmodes, i.e., $TE_2$ and $TE_0$ modes, $TE_1$ and $TE_3$ modes. It should be noted that, due to fabrication errors and interference in the measurement setup, all other modes are also excited. To make sure that all other modes are negligible compared to the expected output modes, AB and CD sections in Fig. 2(a) are shortened to reduce the evolution adiabaticity in designing the final device, guaranteeing that the expected modes are dominantly excited. As a result, the expected modes have significantly larger amplitudes than the modes excited by fabrication imperfections.

## Discussion

Our method to selectively govern the non-adiabaticity degree can be extended to enable chiral switching of a larger number of eigenstates in higher-order non-Hermitian Hamiltonian systems. We have theoretically established an eight-order Hamiltonian, and numerically demonstrated an eight-coupled silicon waveguides system for chiral switching of four pairs of eigenmodes. More details can be found in Supplementary Note 8. In this context, it was recently shown theoretically that the output state can be flexibly selected by winding around exceptional curves by crossing diabolic points in multimode systems[46]. Such a design strategy relies on a complex hybrid coupling mechanism, which may be challenging to realize in practice, especially in a planar integrated photonic system.

In conclusion, we have demonstrated that it is possible to selectively govern the non-adiabaticity degree during the Hamiltonian evolution in encircling EPs, dividing the eigenstate space into multi-state Hamiltonian systems within several direct sum subspaces. The mutual excitation between eigenstates belonging to different subspaces can be avoided by selectively setting a zero non-adiabaticity degree for some transitions. Chiral switching for four eigenstates was theoretically predicted and experimentally demonstrated in four-coupled silicon waveguides. Our approach for multi-state chiral switching is general and extendable to higher-order non-Hermitian Hamiltonian systems. Our results not only provide a fundamental basis for studying multi-state evolution in non-Hermitian systems, but may also be beneficial to develop novel multi-mode photonic devices.

## Materials and methods

**Fabrications.** Such a four-eigenmode chiral converter was fabricated by a combination of three-step electron-beam lithography (EBL), inductively coupled plasma (ICP) etching, electron-beam evaporation (EBE), and plasma-enhanced chemical vapor deposition (PECVD). EBL and EBE were first used to form the Aurum marks on an SOI wafer for alignment. Next, EBL and ICP were employed to define the waveguide pattern and transfer it onto the SOI wafer. Finally, a chromium layer on top of the silicon waveguide was formed by a third-step EBL with careful alignment, EBE and lift-off process, which was followed by PECVD to deposit a 1-μm-thick $SiO_2$ cladding layer covering the entire device. More fabrication details are given in Supplementary Note 6 and Supplementary Fig. S6.

**Measurement.** The near infrared light source is provided by an amplified spontaneous emission (ASE) broadband light source. The polarization of the light source is adjusted by polarization beam splitter (PBS) and polarization controller (PC) before light is coupled into the grating coupler (GC) through the fiber. The emergent light from the SOI chip is coupled back into the fiber through the GC, and reaches 50/50 coupler, which is connected to the optical power meter and spectrometer. The optical power meter is used to adjust the angle between the fiber and the GC so as to maximize the coupling efficiency between them. The spectrometer is used to extract the transmittance for all the output modes. More measurement details are given in Supplementary Note 7 and Supplementary Fig. S7.

## Acknowledgements

This work has been supported by National Natural Science Foundation of China (Grant No. 12074137), National Key Research and Development Project of China (Grant No. 2021YFB2801903), and Science, Technology and Innovation Commission of Shenzhen Municipality (Grant No. JCYJ20220530161010023), Air Force Office of Scientific Research MURI program, and the Simons Foundation. We thank Pan Li in the Center of Micro-Fabrication and Characterization (CMFC) of WNLO for the support in plasma enhanced chemical vapor deposition, and the Center for Nanoscale Characterization & Devices (CNCD), WNLO, HUST for the support in SEM measurement. We thank Ming Deng in the Huazhong University of Science and Technology for redacting the manuscript.

## Author contributions

A.L., and L.C. conceived the idea and initiated the work. A.A. and L.C. guided the project. A.L. and L.C. developed the theoretical framework. A.L. performed the numerical simulations, and performed the measurements. A.A. discussed the results. A.L., A.A., and L.C. wrote the manuscript and all authors reviewed the manuscript.

## Competing interests

The authors declare that they have no competing interests.

## Data availability

All relevant data that support the findings of this study are available from the corresponding author upon reasonable request.

## Code availability

The code that supports the plots within this paper is available from the corresponding authors upon reasonable request.

# References

1. Guo, A. *et al.* Observation of PT-symmetry breaking in complex optical potentials. *Phys. Rev. Lett.* **103**, 093902 (2009).

2. Rüter, C. E. *et al.* Observation of parity–time symmetry in optics. *Nat. Phys.* **6**, 192-195 (2010).

3. Miri, M.-A. & Alù, A. Exceptional points in optics and photonics. *Science* **363**, eaar7709 (2019).

4. Krasnok, A., Nefedkin, N. & Alù, A. Parity-Time Symmetry and Exceptional Points [Electromagnetic Perspectives]. *IEEE Antennas and Propagation Magazine* **63**, 110-121 (2021).

5. Ding, K., Ma, G., Zhang, Z. Q. & Chan, C. T. Experimental Demonstration of an Anisotropic Exceptional Point. *Phys. Rev. Lett.* **121**, 085702 (2018).

6. Tang, W. *et al.* Exceptional nexus with a hybrid topological invariant. *Science* **370**, 1077-1080 (2020).

7. Chen, H.-Z. *et al.* Revealing the missing dimension at an exceptional point. *Nat. Phys.* **16**, 571-578 (2020).

8. Dong, Z., Li, Z., Yang, F., Qiu, C.-W. & Ho, J. S. Sensitive readout of implantable microsensors using a wireless system locked to an exceptional point. *Nat. Electron.* **2**, 335-342 (2019).

9. Shao, L. *et al.* Non-reciprocal transmission of microwave acoustic waves in nonlinear parity–time symmetric resonators. *Nat. Electron.* **3**, 267-272 (2020).

10. Yang, X. *et al.* Observation of Transient Parity-Time Symmetry in Electronic Systems. *Phys. Rev. Lett.* **128**, 065701 (2022).

11. Li, Y. *et al.* Anti–parity-time symmetry in diffusive systems. *Science* **364**, 170-173 (2019).

12. Klauck, F. *et al.* Observation of PT-symmetric quantum interference. *Nat. Photon.* **13**, 883-887 (2019).

13. Liao, Q. *et al.* Experimental Measurement of the Divergent Quantum Metric of an Exceptional Point. *Phys. Rev. Lett.* **127**, 107402 (2021).

14. Liu, W., Wu, Y., Duan, C.-K., Rong, X. & Du, J. Dynamically Encircling an Exceptional Point in a Real Quantum System. *Phys. Rev. Lett.* **126**, 170506 (2021).

15. Miri, M.-A., LiKamWa, P. & Christodoulides, D. N. Large area single-mode parity–time-symmetric laser amplifiers. *Opt. Lett.* **37**, 764-766 (2012).

16. Hodaei, H., Miri, M.-A., Heinrich, M., Christodoulides, D. N. & Khajavikhan, M. Parity-time-

symmetric microring lasers. *Science* **346**, 975-978 (2014).

17. Feng, L., Wong, Z. J., Ma, R.-M., Wang, Y. & Zhang, X. Single-mode laser by parity-time symmetry breaking. *Science* **346**, 972-975 (2014).

18. Hodaei, H. *et al.* Enhanced sensitivity at higher-order exceptional points. *Nature* **548**, 187-191 (2017).

19. Hokmabadi, M. P., Schumer, A., Christodoulides, D. N. & Khajavikhan, M. Non-Hermitian ring laser gyroscopes with enhanced Sagnac sensitivity. *Nature* **576**, 70-74 (2019).

20. Peters, K. J. H. & Rodriguez, S. R. K. Exceptional Precision of a Nonlinear Optical Sensor at a Square-Root Singularity. *Phys. Rev. Lett.* **129**, 013901 (2022).

21. Duggan, R., A. Mann, S. & Alù, A. Limitations of Sensing at an Exceptional Point. *ACS Photonics* **9**, 1554-1566 (2022).

22. Smith, D. D., Chang, H., Mikhailov, E. & Shahriar, S. M. Beyond the Petermann limit: Prospect of increasing sensor precision near exceptional points. *Phys. Rev. A* **106**, 013520 (2022).

23. Zhen, B. *et al.* Spawning rings of exceptional points out of Dirac cones. *Nature* **525**, 354-358 (2015).

24. Zhao, H. *et al.* Non-Hermitian topological light steering. *Science* **365**, 1163-1166 (2019).

25. Xia, S. *et al.* Nonlinear tuning of PT symmetry and non-Hermitian topological states. *Science* **372**, 72-76 (2021).

26. Gilary, I., Mailybaev, A. A. & Moiseyev, N. Time-asymmetric quantum-state-exchange mechanism. *Phys. Rev. A* **88**, 010102 (2013).

27. Milburn, T. J. *et al.* General description of quasiadiabatic dynamical phenomena near exceptional points. *Phys. Rev. A* **92**, 052124 (2015).

28. Doppler, J. *et al.* Dynamically encircling an exceptional point for asymmetric mode switching. *Nature* **537**, 76-79 (2016).

29. Yoon, J. W. *et al.* Time-asymmetric loop around an exceptional point over the full optical communications band. *Nature* **562**, 86-90 (2018).

30. Zhang, X.-L., Jiang, T. & Chan, C. T. Dynamically encircling an exceptional point in anti-parity-time symmetric systems: asymmetric mode switching for symmetry-broken modes. *Light Sci. Appl.* **8**, 88 (2019).

31. Li, A. *et al.* Hamiltonian Hopping for Efficient Chiral Mode Switching in Encircling

Exceptional Points. *Phys. Rev. Lett.* **125**, 187403 (2020).

32. Liu, Q. *et al.* Efficient Mode Transfer on a Compact Silicon Chip by Encircling Moving Exceptional Points. *Phys. Rev. Lett.* **124**, 153903 (2020).

33. Schumer, A. *et al.* Topological modes in a laser cavity through exceptional state transfer. *Science* **375**, 884-888 (2022).

34. Wei, Y. *et al.* Anti-parity-time symmetry enabled on-chip chiral polarizer. *Photonics Res.* **10**, 76-83 (2022).

35. Li, A. *et al.* Riemann-Encircling Exceptional Points for Efficient Asymmetric Polarization-Locked Devices. *Phys. Rev. Lett.* **129**, 127401 (2022).

36. Li, A. *et al.* Exceptional points and non-Hermitian photonics at the nanoscale. *Nat. Nanotechnol.* **18**, 706-720 (2023).

37. Zhang, X.-L. & Chan, C. T. Dynamically encircling exceptional points in a three-mode waveguide system. *Commun. Phys.* **2**, 63 (2019).

38. Gandhi, H. K., Laha, A., Dey, S. & Ghosh, S. Chirality breakdown in the presence of multiple exceptional points and specific mode excitation. *Opt. Lett.* **45**, 1439-1442 (2020).

39. Yu, F., Zhang, X.-L., Tian, Z.-N., Chen, Q.-D. & Sun, H.-B. General Rules Governing the Dynamical Encircling of an Arbitrary Number of Exceptional Points. *Phys. Rev. Lett.* **127**, 253901 (2021).

40. Born, M. & Fock, V. Beweis des Adiabatensatzes. *Z. Phys.* **51**, 165-180 (1928).

41. Kato, T. On the Adiabatic Theorem of Quantum Mechanics. *J. Phys. Soc. Jpn.* **5**, 435-439 (1950).

42. Avron, J. E. & Elgart, A. Adiabatic Theorem without a Gap Condition. *Commun. Math. Phys.* **203**, 445-463 (1999).

43. Ashida, Y., Gong, Z. & Ueda, M. Non-Hermitian physics. *Adv. Phys.* **69**, 249-435 (2020).

44. Huang, W.-P. Coupled-mode theory for optical waveguides: an overview. *Journal of the Optical Society of America A* **11**, 963-983 (1994).

45. Shu, X. *et al.* Fast encirclement of an exceptional point for highly efficient and compact chiral mode converters. *Nat. Commun.* **13**, 2123 (2022).

46. Arkhipov, I. I., Miranowicz, A., Minganti, F., Özdemir, Ş. K. & Nori, F. Dynamically crossing diabolic points while encircling exceptional curves: A programmable symmetric-asymmetric multimode switch. *Nat. Commun.* **14**, 2076 (2023).

# Figures

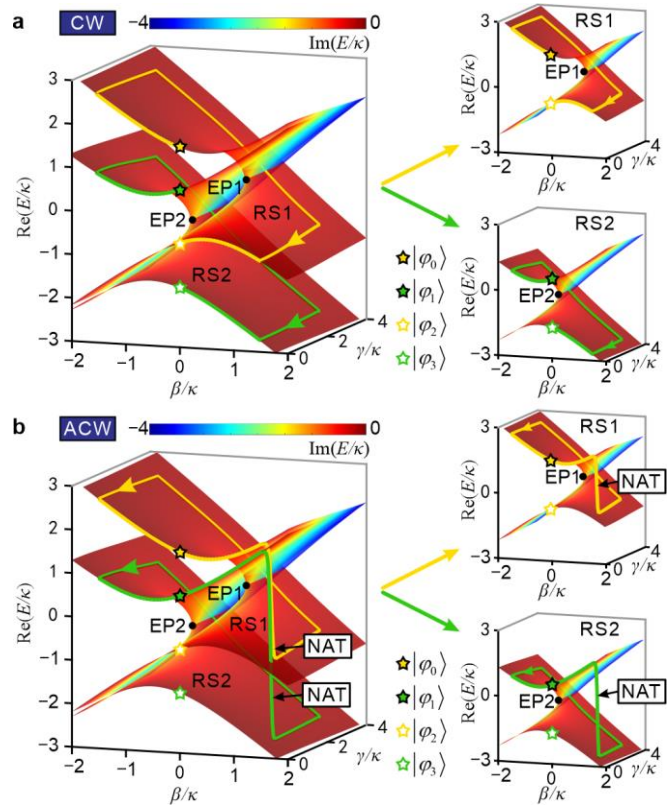

**Fig. 1 System states evolving on the Riemann surfaces. a** CW and **b** ACW loops on the RS formed by the real part of the eigenvalues. The colour of the surface indicates the imaginary part of eigenvalues. The yellow and green lines denote the evolution paths, when $|\varphi_0\rangle$ and $|\varphi_1\rangle$ are injected, respectively. The eigenvalue spectrum is split into two separate RS1 and RS2, as shown in the right panels.

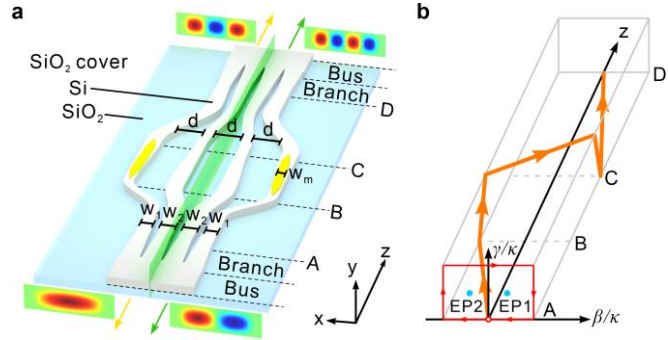

**Fig. 2 Silicon waveguides for four-state chiral switching. a** Four coupled silicon waveguides to demonstrate four-state chiral switching. The device is symmetric with respect to the green plane. The silicon waveguides have a height of 220 nm on a silicon-on-isolator (SOI) wafer, and are covered by a 1-μm-thick SiO₂ layer in order to provide structural protection. The first (second) and fourth (third) waveguides have the same width $w_1$ ($w_2$). The gap separations between all adjacent waveguides, denoted as $d$, are kept constant. The width of the chromium layer is marked by $w_m$. The four coupled waveguides are connected with bus waveguides using branches. **b** Encircling loop around two EPs. The orange line with arrows shows the evolution trajectory, whereas its projection onto $(\beta/\kappa, \gamma/\kappa)$ plane is marked by the redline. The blue lines show the location of two EPs.

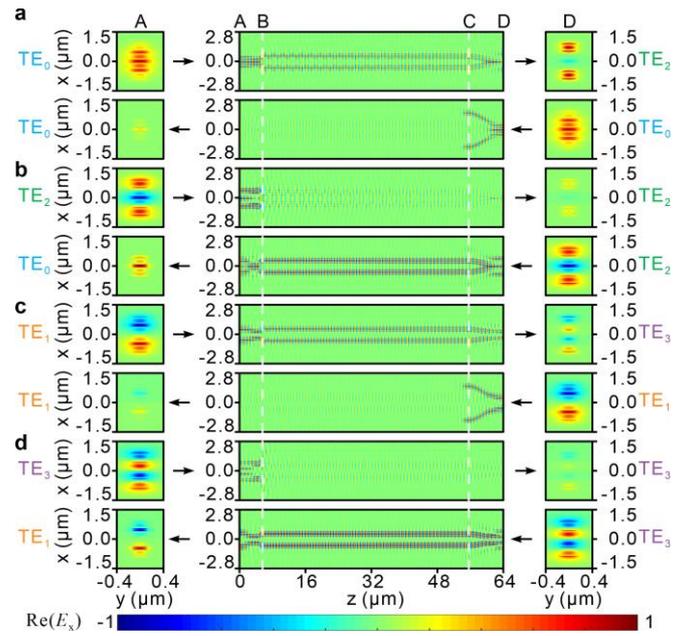

**Fig. 3 Simulated field distributions. a-d** The transverse components of electric field $\mathrm{Re}(E_x)$ for $TE_0$ (**a**), $TE_2$ (**b**), $TE_1$ (**c**), and $TE_3$ (**d**) modes input at 1550 nm. The left and right panels correspond to the cross-sectional field distributions at the two ports. The first and second rows of each graph indicate light injected from left and right ports, respectively.

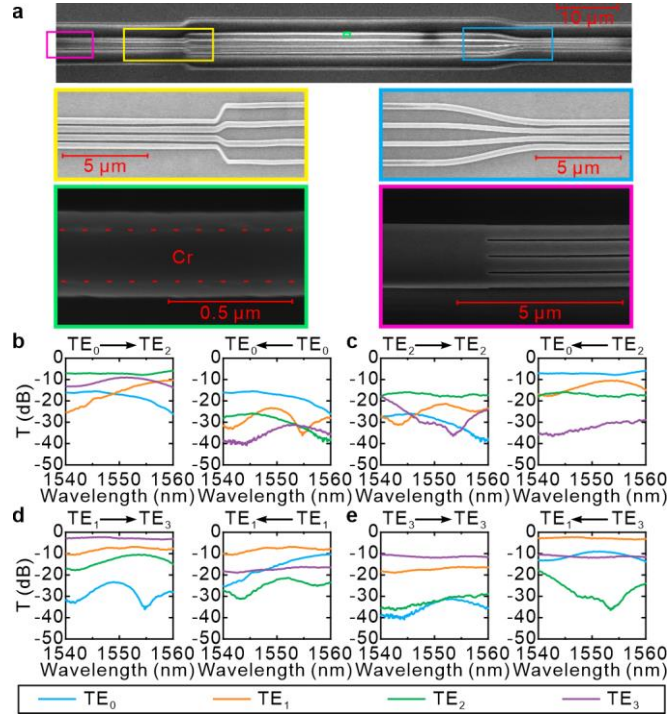

**Fig. 4 Experimental demonstration. a** SEM image of the four-eigenmode chiral switching device. The top panel shows the entire device, where the zoom-in SEM images bounded by the four rectangles are presented in the four panels below. The yellow and blue boxes show the AB and CD regions, respectively, and the pink box shows the branch that connects the bus waveguide and four-coupled waveguides. The green box represents the chromium layer, where the red dash lines indicate the interface between the chromium and silicon. **b-d** Measured transmittance spectra over the wavelength range of 1540-1560 nm, where $TE_0$ (**b**), $TE_2$ (**c**), $TE_1$ (**d**) and $TE_3$ (**e**) modes are injected. The first and second columns in each figure indicate light is injected from left and right ports, respectively.